\documentclass[final,5p,times,twocolumn]{elsarticleMY}

\usepackage{graphicx}
\usepackage{amssymb}
\usepackage{hyperref}

\begin{document}

\begin{frontmatter}

\title{On particle collisions in the gravitational field of
the Kerr black hole}

\author{A.A. Grib\fnref{label1,label2}}
\ead{andrei\_grib@mail.ru}

\author{Yu.V. Pavlov\fnref{label1,label3}}
\ead{yuri.pavlov@mail.ru}

\address[label1]{A. Friedmann Laboratory for Theoretical Physics,
30/32 Griboedov can., St.\,Petersburg 191023, Russia}

\address[label2]{Theoretical Physics and Astronomy Department,
The Herzen  University, 48, Moika, St.\,Petersburg 191186, Russia}

\address[label3]{Institute of Mechanical Engineering, Russian Academy
of Sciences, 61 Bolshoy, V.O., St.\,Petersburg 199178, Russia}

\begin{abstract}
    Scattering of particles in the gravitational field
of Kerr black holes is considered.
    It is shown that scattering energy of particles in the centre of mass
system can obtain very large values not only for extremal black holes
but also for nonextremal ones existing in Nature.
    This can be used for explanation of still unresolved problem of the origin
of ultrahigh energy cosmic rays observed in Auger experiment.
    Extraction of energy after the collision is investigated.
    It is shown that due to the Penrose process the energy of the particle
escaping the hole at infinity can be large.
    Contradictions in the problem of getting high energetic particles
escaping the black hole are resolved.
\end{abstract}


\end{frontmatter}

\section{Introduction}
\label{Intro}

    In~\cite{GribPavlov2007AGN} we put the hypothesis that
Active Galactic Nuclei (AGN) can be the sources of ultrahigh energy particles
in cosmic rays observed recently by the AUGER group (see~\cite{Auger07})
due to the processes of converting dark matter formed by superheavy neutral
particles into visible particles --- quarks, leptons (neutrinos), photons.
    Such processes as it was discussed previously
in~\cite{GrPv1,GribPavlov2008KLGN} could lead to the
origination of visible matter from the dark matter particles in the
early Universe when the energy of particles was of the Grand
Unification (GU) order ($10^{13} - 10^{14}$\,GeV) i.e. at the end of
inflation era.

    If AGN are rotating black holes then in~\cite{GribPavlov2007AGN}
we discussed the idea that ``This black hole acts as a cosmic
supercollider in which superheavy particles of dark matter are
accelerated close to the horizon to the GU energies and can be
scattering in collisions.''
    It was also shown that in Penrose process~\cite{Penrose69}
dark matter particle can decay on two particles, one with the
negative energy, the other with the positive one and particles of
very high energy of the GU order can escape the black hole.
    Then these particles due to interaction with photons close to
the black hole will loose energy analogously up to the
Greisen-Zatsepin-Kuzmin limit in cosmology~\cite{GZK}.

    Processes of converting dark matter particles into visible ones
as well as collisions of dark matter particles close to the horizon
of the black hole in the AGN can play important role in the solution
of the problem of cusps as it was discussed in~\cite{DEFriedmann09}.
    According to the hierarchical model the galaxies were formed due to
initial inhomogeneities of dark matter distribution.
    But after origination of rotating black holes at the centre of future
galaxies large part of the dark matter became redistributed so that
it can be mainly present in the halo of galaxy~\cite{DEFriedmann09}.

    The problem of the origin of ultrahigh energy cosmic rays (UHECR)
is a major unsolved issue in astroparticle physics.
    Surely there can be different mechanisms of getting UHECR from AGN.
    Our aim is full investigation of the role of processes of
collisions and decays of elementary particles close to the horizon of
the Kerr black holes in the AGN.
    These particles can be superheavy particles of dark matter or
usual protons, iron nuclei, etc.
    In the latter cases for usual particles to get large energy one must
consider multiple scattering.

    First calculations of the scattering of particles in the ergosphere
of the rotating black hole, taking into account the Penrose process, with
the result that particles with high energy can escape the black hole, were
made in~\cite{PiranShahamKatz75}.
    Recently in~\cite{BanadosSilkWest09} it was shown that for the rotating
black hole (if it is close to the critical one) the energy of scattering
is unlimited.
    The result of~\cite{BanadosSilkWest09} was criticized
in~\cite{BertiCardosoGPS09,JacobsonSotiriou09}
where the authors claimed that
if the black hole is not a critical rotating black hole so that its
dimensionless angular momentum $A \ne 1$ but $ A=0.998$ then
the energy is limited.

    In this paper we show that the energy of scattering in the centre of mass
system can be still unlimited in the cases of multiple scattering.
    In Section~\hyperref[2secBHColl]{2}, we calculate this energy, reproduce
the results of~\cite{BanadosSilkWest09}--\cite{JacobsonSotiriou09}
and show that in some cases (multiple scattering) the results
of~\cite{BertiCardosoGPS09,JacobsonSotiriou09} on the limitations of
the scattering energy for nonextremal black holes are not valid.
    In Section~\hyperref[31subsecEnUFN]{2.1} an evaluation of the time
necessary for the freely falling particle to get the unbounded large energy
in collision with the other particle in the centre of mass system is made.
    It is proved that in order to have infinitely large energy of collision
on the horizon infinitely large time from the beginning of free falling of
the particle on the rotating black hole is needed.
    Particle collisions inside the rotating black hole are
considered in Section~\hyperref[3secEnUFN]{3}.
    In Section~\hyperref[3secBHColl]{4}, we obtain the results for the extraction
of the energy after collision in the field of the Kerr's metric.
    It occurs that the Penrose process plays important role for getting
larger energies of particles at infinity.
    Our calculations show that the conclusion of~\cite{JacobsonSotiriou09}
about the impossibility of getting at infinity the energy larger
than the initial one in particle collisions close to the black hole is wrong.

    The system of units $G=c=1$ is used in the paper.

{\section{The energy of collision in the field of black holes}
\label{2secBHColl}
}

    Let us consider particles falling on the rotating chargeless black hole.
    The Kerr's metric of the rotating black hole in Boyer--Lindquist
coordinates has the form                    
    \begin{eqnarray}
d s^2 = d t^2 -
\frac{2 M r \, ( d t - a \sin^2 \! \theta\, d \varphi )^2}{r^2 + a^2 \cos^2
\! \theta } \hspace{33pt}
\nonumber \\
- (r^2 + a^2 \cos^2 \! \theta ) \left( \frac{d r^2}{\Delta} + d \theta^2 \right)
- (r^2 + a^2) \sin^2 \! \theta\, d \varphi^2, \ \
\label{Kerr}
\end{eqnarray}
    where
    \begin{equation} \label{Delta}
\Delta = r^2 - 2 M r + a^2,
\end{equation}
    $M$ is the mass of the black hole, $J=aM$ is angular momentum.
    In the case $a=0$ the metric~(\ref{Kerr}) describes the static chargeless
black hole in Schwarzschild coordinates.
    The event horizon for the Kerr's black hole corresponds to the value
    \begin{equation}
r = r_H \equiv M + \sqrt{M^2 - a^2} .
\label{Hor}
\end{equation}
    The Cauchy horizon is
    \begin{equation}
r = r_C \equiv M - \sqrt{M^2 - a^2} . \label{HorCau}
\end{equation}
    The surface called ``the static limit'' is defined by the expression
     \begin{equation}
r = r_0 \equiv M + \sqrt{M^2 - a^2 \cos^2 \! \theta} \,.
\label{PrSt}
\end{equation}
    The region of space-time between the horizon and the static limit is
ergosphere.

    For equatorial ($\theta=\pi/2$) geodesics in Kerr's metric~(\ref{Kerr}) one
obtains (\cite{Chandrasekhar}, \S\,61)
    \begin{equation} \label{geodKerr1}
\frac{d t}{d \tau} = \frac{1}{\Delta} \left[ \left(
r^2 + a^2 + \frac{2 M a^2}{r} \right) \varepsilon - \frac{2 M a}{r} L \, \right],
\end{equation}
    \begin{equation}
\frac{d \varphi}{d \tau} = \frac{1}{\Delta} \left[ \frac{2 M a}{r}\,
\varepsilon + \left( 1 - \frac{2 M}{r} \right) L \, \right],
\label{geodKerr2}
\end{equation}
    \begin{equation} \label{geodKerr3}
\left( \frac{d r}{d \tau} \right)^2 = \varepsilon^2 +
\frac{2 M}{r^3} \, (a \varepsilon - L)^2 +
\frac{a^2 \varepsilon^2 - L^2}{r^2} - \frac{\Delta}{r^2}\, \delta_1 \,,
\end{equation}
    where
    $\delta_1 = 1 $ for timelike geodesics
($\delta_1 = 0 $ for isotropic geodesics),
$\tau$ is the proper time of the moving particle,
$\varepsilon={\rm const} $ is the specific energy:
the particle with rest mass~$m$ has the energy $\varepsilon m $ in the
gravitational field~(\ref{Kerr});
$ L m = {\rm const} $ is the angular momentum of the particle relative
to the axis orthogonal to the plane of movement.

    Let us find the energy $E_{\rm c.m.}$ in the centre of mass system
of two colliding particles with the same rest~$m$ in arbitrary gravitational
field.
    It can be obtained from
    \begin{equation} \label{SCM}
\left( E_{\rm c.m.}, 0\,,0\,,0\, \right) =
m u^i_{(1)} + m u^i_{(2)}\,,
\end{equation}
    where $u^i=dx^i/ds$.
    Taking the squared~(\ref{SCM}) and due to $u^i u_i=1$ one obtains
    \begin{equation} \label{SCM2}
E_{\rm c.m.} = m \sqrt{2}\, \sqrt{1+ u_{(1)}^i u_{(2) i}} \,.
\end{equation}
    The scalar product does not depend on the choice of the coordinate frame
so~(\ref{SCM2}) is valid in an arbitrary coordinate system and for arbitrary
gravitational field.

    We denote~$x=r/M$, \ $ x_H=r_H/M$, \ $ x_C=r_C/M$, \ $ A=a/M$,
\ $ l_n=L_n/M$.
    Apply the formula~(\ref{SCM2}) for calculation of the energy in the centre
of mass frame of two colliding particles with angular momenta $L_1, \, L_2$,
which are nonrelativistic at infinity ($\varepsilon_1 = \varepsilon_2 = 1$)
and are moving in Kerr's metric.  Using~(\ref{Kerr}),
(\ref{geodKerr1})--(\ref{geodKerr3}) one obtains~\cite{BanadosSilkWest09}
    \begin{eqnarray}
\frac{E_{\rm c.m.}^2}{2\, m^2} =
\frac{1}{x \Delta_x} \biggl[ 2 x^2 (x-1) + l_1 l_2 (2-x)
\hspace{22pt}
\nonumber \\
+\, 2 A^2 (x+1) - 2 A (l_1 +l_2 ) \hspace{49pt}
 \label{KerrL1L2} \\
\hspace*{-4pt}
-\sqrt{\left( 2 x^2 + 2 (l_1 \!-\! A)^2 -l_1^2 x \right)
\left(2 x^2 + 2 (l_2 \!-\! A)^2 -l_2^2 x \right) } \biggr],
\nonumber
\end{eqnarray}
    where
    \begin{equation} \label{DxxHxC}
\Delta_x = x^2 - 2 x + A^2 = (x-x_H) (x-x_C) \,.
\end{equation}

    To find the limit $r \to r_H$ for the black hole with a given angular
momentum~$A$ one must take in~(\ref{KerrL1L2}) $x = x_H + \alpha$
with $\alpha \to 0 $ and do calculations up to the order $\alpha^2$.
    Taking into account $ A^2 = x_H x_C$, $x_H + x_C=2$, after resolution
of uncertainties in the limit $\alpha \to 0 $ one obtains
    \begin{equation} \label{KerrLimA}
\frac{E_{\rm c.m.}(r \to r_H) }{2 m}
= \sqrt{1 + \frac{(l_1-l_2)^2}{2 x_C (l_1 -l_H) (l_2 - l_H)}} \,,
\end{equation}
    where
    \begin{equation} \label{KerrlH}
l_H = \frac{2 x_H}{A} =  \frac{2}{A} \left( 1 + \sqrt{1-A^2} \, \right).
\end{equation}

    In the limit $r \to r_H$ for the extremal black hole one obtains
the expression
    \begin{equation} \label{KerrL1L2lim}
a=M \ \Rightarrow  \ E_{\rm c.m.}(r \to r_H) = \sqrt{2}\, m
\sqrt{ \frac{l_2-2}{l_1 - 2} + \frac{l_1-2}{l_2 - 2} } \,,
\end{equation}
    given first in~\cite{BanadosSilkWest09},
showing the unlimited increasing of the energy of collision when
the specific angular momentum of one of falling particles goes to the
limiting possible value equal to $2 M$
necessary for achieving the horizon of the extremal black hole.

    Formula~(\ref{geodKerr3}) leads to limitations on the possible values
of the angular momentum of falling particles:
the massive particle free falling in the black hole with dimensionless
angular momentum~$A$ being nonrelativistic at infinity ($\varepsilon = 1 $)
to achieve the horizon of the black hole must have angular momentum
from the interval
    \begin{equation} \label{geodKerr5}
- 2 \left( 1 + \sqrt{1+A}\, \right) =l_L \le l
\le l_R = 2 \left( 1 + \sqrt{1-A}\, \right).
\end{equation}
    Note that for the mentioned limiting values the right hand side of
the formula~(\ref{geodKerr3}) is zero for
    \begin{equation} \label{geodKerrxRL}
x_R = 2 \left( 1+ \sqrt{1-A}\, \right) - A ,
\ \ \
x_L=2 \left( 1+ \sqrt{1+A}\, \right) + A \ \
\end{equation}
    for $ l=l_R$ and $l=l_L$ correspondingly.

\begin{figure}[ht]
\centering
\includegraphics[height=54mm]{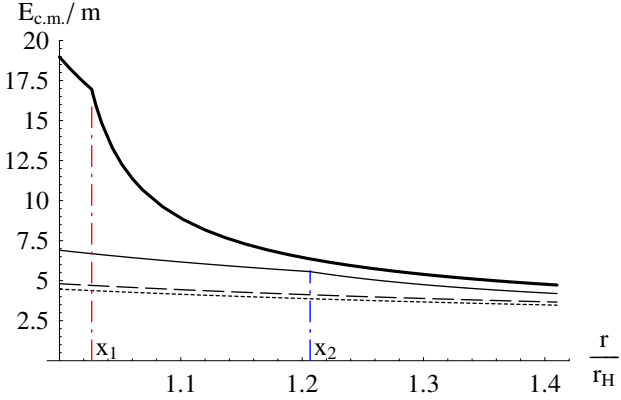}
\caption{The dependence of the energy of particle collision on
the coordinate~$r$.}
\label{EmaxCMr}
\end{figure}

     The dependence of the energy of collision $E_{\rm c.m.}$ on the radial
coordinate for $l_1 = l_R, \, l_2=l_L$
(see~(\ref{geodKerr5})) is given on Fig.~\ref{EmaxCMr}.
    The boldface line corresponds to~$A=0.998$,
the ordinary curve to~$A=0.9$, the dotted to~$A=0.5$,
the line formed by points describes nonrotating black hole~($A=0$).
    The fractures on dense line, denoted by dotted lines correspond
to values $x=x_R(A) M /r_H(A)$ (see $x_R$ in~(\ref{geodKerrxRL}))
i.e. to zeroes of the expressions in roots in formula~(\ref{KerrL1L2}).

\begin{figure}[ht]
\centering
\includegraphics[height=52mm]{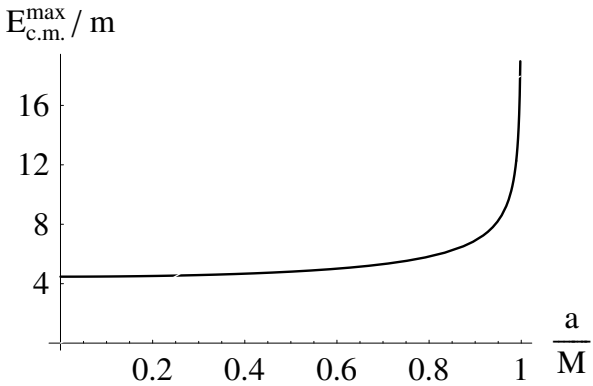}
\caption{The dependence of the maximal energy of collision for particles
falling from infinity on the black hole angular momentum.}
\label{EmaxCM}
\end{figure}

    Putting the limiting values of angular momenta $l_L, l_R$
into the formula~(\ref{KerrLimA}) one obtains the maximal values
of the energy of collision for particles falling from infinity
    \begin{eqnarray} \label{KerrLimAMax}
E_{\rm c.m.}^{\, \rm max}(r \to r_H) =
    \hspace{22pt} \nonumber \\
= \frac{2m}{\sqrt[4]{1-A^2}} \,
\sqrt{\frac{1-A^2+\left( 1+ \sqrt{1+A} +\sqrt{1-A} \right)^2}
{1+\sqrt{1-A^2}} }\,.
\end{eqnarray}
    The dependence of $E_{\rm c.m.}^{\, \rm max}$ on the angular
momentum of the black hole is shown on Fig.~\ref{EmaxCM}.

    For nonrotating black hole the maximal energy of collision due
to~(\ref{KerrLimAMax}) is $E_{\rm c.m.}^{\, \rm max}/ m = 2 \sqrt{5} $
(first found in~\cite{Baushev09}).

    For $A=1-\epsilon$ with $\epsilon \to 0$ formula~(\ref{KerrLimAMax})
gives:
    \begin{equation} \label{KerrimAE}
E_{\rm c.m.}^{\, \rm max}(r \to r_H)
\sim 2 \left( 2^{1/4}+2^{-1/4} \right) \frac{m}{\epsilon^{1/4}}
\approx \frac{m \cdot 4,06}{\epsilon^{1/4} } \,.
\end{equation}
    (see~(8) in~\cite{JacobsonSotiriou09}).
    So even for values close to the extremal $A=1$ of the rotating black hole
$E_{\rm c.m.}^{\, \rm max}/ m$ can be not very large as it is mentioned
in~\cite{BertiCardosoGPS09,JacobsonSotiriou09}.
    So for $A_{\rm max} =0.998 $ considered as the maximal possible
dimensionless angular momentum of the astrophysical black holes
(see~\cite{Thorne74}), from~(\ref{KerrLimAMax}) one obtains
$ E_{\rm c.m.}^{\, \rm max} /m \approx 18.97 $.

    Does it mean that in real processes of particle scattering in
the vicinity of the rotating nonextremal black holes the scattering energy
is limited so that no Grand Unification or even Planckean energies can
be obtained?
    Let us show that the answer is no!
    If one takes into account the possibility of multiple scattering so that
the particle falling from the infinity on the black hole with some fixed
angular momentum changes its momentum in the result of interaction with
particles in the accreting disc and after this is again scattering close to
the horizon then the scattering energy can be unlimited.

    The limiting value of the angular momentum of the particle close
to the horizon of the black hole can be obtained from the condition
of positive derivative in~(\ref{geodKerr1}) $dt /d \tau > 0$,
i.e. going ``forward'' in time.
    So close to the horizon one has the condition
$l < \varepsilon 2 x_H/A$ which for $ \varepsilon =1$
gives the limiting value $l_H$.

    From~(\ref{geodKerr3}) one can obtain the permitted interval in~$r$ for
particles with $ \varepsilon = 1 $ and angular momentum $l = l_H - \delta $.
    In the second order in~$\delta$ close to the horizon one obtains
    \begin{equation} \label{KerrIntl}
l = l_H - \delta \ \ \Rightarrow \ \ \
x \lesssim x_H + \frac{\delta^2 x_C^2}{4 x_H \sqrt{1-A^2} } \,.
\end{equation}
    If the particle falling from the infinity with $ l \le l_R$ arrives
to the region defined by~(\ref{KerrIntl}) and here it interacts with other
particles of the accretion disc or it decays on more light particle so
that it gets the larger angular momentum $l_1 = l_H - \delta $,
then due to~(\ref{KerrLimA}) the scattering energy in the centre of mass
system is
    \begin{equation} \label{KerrInEn}
E_{\rm c.m.} \approx \frac{m}{\sqrt{\delta}} \,
\sqrt{ \frac{ 2(l_H - l_2) }{ 1- \sqrt{1 - A^2} } }
\end{equation}
    and it increases without limit for $\delta \to 0$.
    For $A_{\rm max} =0.998 $ and $l_2=l_L$, \
$ E_{\rm c.m.} \approx 3.85 m / \sqrt{\delta} $.  

    Note that for rapidly rotating black holes $A= 1 - \epsilon$
the difference between $l_H$ and $l_R$ is not large
    \begin{eqnarray}
l_H - l_R &=& 2 \frac{\sqrt{1-A}}{A} \left( \sqrt{1-A} + \sqrt{1+A} -A \right)
\nonumber \\
&\approx& 2 (\sqrt{2}-1) \sqrt{\epsilon}\,, \ \ \ \epsilon \to 0 \,.
\label{KerrInDLR}
\end{eqnarray}
    For $A_{\rm max} =0.998 $, \ $ l_H - l_R \approx 0.04$
so the possibility of getting small additional angular momentum in
interaction close to the horizon seems much probable.

    That is why we come to a conclusion that unboundedly large energy can be
obtained in the centre of mass system of scattering particles.
    This means that physical processes not only on the Grand Unification
scale but even on the Planckean energy can go in this case.

    One can obtain the same result of getting unbounded large energy in
the centre of mass frame for particles with different masses $m_1 \ne m_2$
and $\varepsilon_1 \ne \varepsilon_2$~\cite{GrPv10c}.

    However an important question is to look how particles with large energy
can be extracted from the black hole?
    Some energy surely will be loosed due to the red shift, but as
we'll show in this paper ultrarelativistic particles still can be observed
outside the black hole.
    Observation of them can give us insight into the Grand Unification
and Planckean physics near the horizon.

    Before answering on this question let us investigate the problem about the time
needed for the energy to grow unboundedly large.

{\subsection{The time of movement before the collision with unbounded energy}
\label{31subsecEnUFN}
}

    Let us show that in order to get the unboundedly growing energy one must
have the time interval from the beginning of the falling inside the black
hole to the moment of collision also growing infinitely.
    This is connected with the fact of infinity of coordinate interval of
time needed for a freely particle to cross the horizon of the black hole
(for Schwarzschild metric this question was considered
in~\cite{GribPavlov2008UFN}).

    From Fig.~\ref{EmaxCMr} one can see that the collision energy can get
large values in the centre of mass system only if the collision occurs close
to the horizon.
    Unboundedly large energy of collisions outside of a black hole
are possible only for collisions on horizon $x \to x_H$
(see~(\ref{KerrL1L2}), (\ref{KerrLimA})).

    From equation of the equatorial geodesic~(\ref{geodKerr1}),
(\ref{geodKerr3}) for a particle with dimensionless angular momentum~$l$
and specific energy $\varepsilon=1$ (i.e. the particle is non relativistic
at infinity) falling on the black hole with dimensionless angular momentum~$A$
one obtains
    \begin{equation} \label{Kerrdrdt}
\frac{d r}{d t} =  - \frac{(x- x_H) (x- x_C)}{\sqrt{x}} \,
\frac{\sqrt{ 2 x^2 - l^2 x + 2 (A -l)^2} }{x^3 +A^2 x + 2 A (A-l) }\,.
\end{equation}
    So the coordinate time (proper time of the observer at rest far from
the black hole) of the particle falling from some point $ r_0 = x_0 M $
to the point $r_f = x_f M > r_H$ is equal to
    \begin{equation} \label{KerrDelt}
\Delta t = M \int \limits_{x_f}^{x_0}
\frac{\sqrt{x} \left(x^3 +A^2 x + 2 A (A-l) \right)\, d x}{(x- x_H) (x- x_C)
\sqrt{ 2 x^2 - l^2 x + 2 (A -l)^2}} \,.
\end{equation}

    In case of the extremal rotating black hole ($A=1$, $x_R =x_H =1 $)
and the limiting value of the angular momentum $l=2$
the integral~(\ref{KerrDelt}) is equal to
    \begin{equation} \label{KerrDeltA1l2}
\Delta t = \frac{M}{\sqrt{2}} \left. \left(
\frac{2 \sqrt{x}\, (x^2 + 8 x - 15) }{ 3(x-1)} + 5 \ln
\frac{\sqrt{x} - 1}{\sqrt{x} + 1} \, \right)\, \right|_{x_f}^{x_0}
\end{equation}
    and it diverges as $ (x_f - 1)^{-1} $ for $ x_f \to 1 $.

    In all other cases, when  $A \le 1$ and angular moment $l <l_H $,
integral (\ref{KerrDelt}) diverges logarithmically if $ x_f \to x_H $.
    This follows from equality
    \begin{equation} \label{KerrZnam}
\hspace*{-17pt}
x^3 + A^2 x + 2 A ( A-l) =\! (x- x_H) ( x^2\! + x_H x+ 2 x_H) + 2 A ( l_H-l). \
\end{equation}
    Note that for the particle with $\varepsilon=1$ outside the horizon
for $A<1$ the angular momentum $l < l_H $ (see~(\ref{KerrIntl})).
    So for all possible values of $l$ and $A$ to get the collision with
infinitely growing energy in the centre of mass system needs infinitely
large time.

    For the interval of proper time of the free falling to the black hole
particle one obtains from~(\ref{geodKerr3})
    \begin{equation} \label{KerrDeltau}
\Delta \tau = M \int \limits_{x_f}^{x_0}
\frac{\sqrt{x^3}\, d x}{
\sqrt{ 2 x^2 - l^2 x + 2 (A -l)^2}} \,.
\end{equation}
    If the angular momentum of the particle falling inside the black
hole is such that $l_L < l < l_R$ then the proper time is finite
for $ x_f \to x_H $.
    For $A=1$, $l=2$ the integral~(\ref{KerrDeltau})) is equal to
    \begin{equation} \label{KerrDeltauA1l2}
\Delta \tau = \frac{M}{3 \sqrt{2}} \left. \left(
2 \sqrt{x}\, (3+ x) + 3 \ln \frac{\sqrt{x} - 1}{\sqrt{x} + 1}
\, \right)\, \right|_{x_f}^{x_0}
\end{equation}
    and it diverges logarithmically when $ x_f \to 1 $.
    So to get the collision with infinite energy one needs the infinite
interval of as coordinate as proper time of the free falling particle.

    To get large finite energy one must have some finite time~\cite{GrPv10c}.
    So to have the collision of two protons with the energy of the order
of the Grand Unification one must wait for the extremal black hole of
the star mass the time $\sim 10^{24}$\,s,
which is larger than the age of the Universe $\approx 10^{18}$\,s.
    However for the collision with the energy $10^3$ larger than
 that of the LHC one must wait only~$\approx 10^{8}$\,s.

    From~(\ref{geodKerr2}), (\ref{geodKerr3}) for the angle of the particle
falling in equatorial plane of the black hole one obtains
    \begin{equation} \label{KerrDelPhi}
\Delta \varphi = \int \limits_{x_f}^{x_0}
\frac{\sqrt{x} \left(x l + 2 (A-l) \right)\, d x}{(x- x_H) (x- x_C)
\sqrt{ 2 x^2 - l^2 x + 2 (A -l)^2}} \,.
\end{equation}
    If $A \ne 0$, then integral~(\ref{KerrDelPhi}) is divergent
for $ x_f \to x_H $.
    So before collision with infinitely large energy the particle must
commit infinitely large number of rotations around the black hole.

{\section{Collision of particles inside the rotating black hole}
\label{3secEnUFN}
}

    As one can see from formula~(\ref{KerrL1L2}) the infinite value
of the collision energy in the centre of mass system can be obtained inside
the horizon of the black hole on the Cauchy horizon~(\ref{HorCau}).
    Indeed, from~(\ref{DxxHxC}) the zeroes of the denominator
in~(\ref{KerrL1L2}): $ x=x_H, \ x=x_C, \ x=0$.

    Let us find the expression for the collision energy for $x \to x_C$.
    Note that the Cauchy horizon can be crossed by the free falling from
the infinity particle under the same conditions on the angular
momentum~(\ref{geodKerr5}) as in case of the event horizon.
    Denote
    \begin{equation} \label{KerrlC}
l_C = \frac{2 x_C}{A} =  \frac{2}{A} \left( 1 - \sqrt{1-A^2} \, \right).
\end{equation}
    Note that $ l_L < l_C \le l_R \le l_H$.

    To find the limit $r \to r_C$ for the black hole with a given angular
momentum~$A$ one must take in~(\ref{KerrL1L2}) $x = x_C + \alpha$
and do calculations with $\alpha \to 0 $ .
    The limiting energy has two different expressions depending on the values
of angular momenta.
    If
    \begin{equation} \label{lvnu1}
(l_1 - l_C) (l_2 - l_C) > 0 \,,
\end{equation}
i.e. $ l_1, l_2 $ are either both larger than $l_C$, or both smaller than $l_C$,
then
    \begin{equation} \label{EcmVnu1}
\frac{E_{\rm c.m.} (r \to r_C) }{2\, m} = \sqrt{
1 + \frac{(l_1 - l_2)^2}{2 x_H (l_1 - l_C) (l_2 - l_C)} } \,.
\end{equation}
    This formula is similar to~(\ref{KerrLimA})
if everywhere $H \leftrightarrow C$. \
    If
    \begin{equation} \label{lvnu2}
(l_1 - l_C) (l_2 - l_C) < 0 \,,
\end{equation}
    i.e. $ l_2 \in (l_L, \, l_C ), \ \ l_1 \in (l_C, \, l_R )$
(or the opposite), then
    \begin{equation} \label{EcmVnu2}
\frac{E_{\rm c.m.}}{m} = \frac{A}{1 - \sqrt{1-A^2}} \
\sqrt{\frac{2 (l_1 - l_C) (l_C - l_2)}{\sqrt{1-A^2}\, (x - x_C)}}, \  \
x \to x_C. \
\end{equation}
    It is seen that the limit is infinite for all values of angular
momenta $l_1,l_2$~(\ref{lvnu2}).
    This could be interpreted~\cite{Lake10} as the known instability of
the internal Kerr's solution (see~\cite[chap. 14]{NovikovFrolov}).
    However, from Eq.~(\ref{geodKerr1}) we can see
    \begin{equation} \label{EcmVnuLimll}
\frac{d t}{d \tau}(r \to r_C + 0 ) = \left\{
\begin{array}{ll}
+ \infty, & \ {\rm if} \ \ l > l_C\,, \\
- \infty, & \ {\rm if} \ \ l < l_C\,.
\end{array} \right.
\end{equation}
    That is why the collisions with infinite energy can not be
realized (see also~\cite{Lake10}).

{\section{The extraction of energy after the collision in
Schwarzschild's and Kerr's metrics}
\label{3secBHColl}
}

    Let us consider the case when interaction between particles with
masses~$m$, specific energies $\varepsilon_{1}$, $ \varepsilon_{2} $,
specific angular momenta $ L_{1}, \ L_{2} $ falling into a black hole
occurred for some $r>r_H$.
    Let two new particles with rest masses~$\mu$ appeared,
one of them~(1) moved outside the black hole, the other~(2)
moved inside it.
    Denote the specific energies of new particles as
$\varepsilon_{1 \mu}$, $ \varepsilon_{2 \mu} $, their angular momenta
(in units of $\mu$) as $ L_{1 \mu}, \ L_{2 \mu} $, \ $v^i = dx^i/ds $
 --- their 4-velocities.
    Consider particle movement in the equatorial plane of the rotating
black hole.

    Conservation laws in inelastic particle collisions for the energy
and momentum lead to
    \begin{equation} \label{I1}
m ( u_{(1)}^i + u_{(2)}^i ) = \mu ( v_{(1)}^i + v_{(2)}^i ) \,.
\end{equation}
    Equations~(\ref{I1}) for $t$ and $\varphi$-components can be written as
    \begin{equation} \label{I5K}
m (\varepsilon_1 + \varepsilon_2) =
\mu (\varepsilon_{1 \mu} + \varepsilon_{2 \mu}) \,,
\end{equation}
    \begin{equation} \label{I5KL}
m (L_1 + L_2) = \mu (L_{1 \mu} + L_{2 \mu}) \,,
\end{equation}
    i.e. the sum of energies and angular momenta of colliding particles is
conserved in the field of Kerr's black hole.

    For the $r$-component from~(\ref{geodKerr3}) one obtains
    \begin{eqnarray}
- m \left[ \sqrt{ \varepsilon_1^2
+\frac{2 M}{r^3} \,(a \varepsilon_1 - L_1)^2 +
\frac{a^2 \varepsilon_1^2 - L_1^2}{r^2} - \frac{\Delta}{r^2}} \right.
\nonumber  \\[2pt]
+\left. \sqrt{ \varepsilon_2^2 +
\frac{2 M}{r^3} \, (a \varepsilon_2 - L_2)^2 +
\frac{a^2 \varepsilon_2^2 - L_2^2}{r^2} - \frac{\Delta}{r^2}}
\, \right]
    \nonumber \\[2pt]
= \mu \left[ \sqrt{ \varepsilon_{1 \mu}^2
+\frac{2 M}{r^3} \,(a \varepsilon_{1 \mu} - L_{1 \mu})^2 +
\frac{a^2 \varepsilon_{1 \mu}^2 - L_{1 \mu}^2}{r^2} - \frac{\Delta}{r^2}}
\right.
\nonumber  \\[2pt]
-\left. \sqrt{ \varepsilon_{2 \mu}^2 +
\frac{2 M}{r^3} \, (a \varepsilon_{2 \mu} - L_{2 \mu})^2 +
\frac{a^2 \varepsilon_{2 \mu}^2 - L_{2 \mu}^2}{r^2} - \frac{\Delta}{r^2}}
\, \right].
\label{I4K}
\end{eqnarray}
    The signs in~(\ref{I4K}) are put so that the initial particles and
the particle~(2) go inside the black hole while particle~(1) goes
outside the black hole.
    The values $ \varepsilon_{1 \mu}, \varepsilon_{2 \mu}$ are constants
on geodesics (\cite{Chandrasekhar}, \S\,61)
so the problem of evaluation of the energy at infinity extracted from
the black hole in collision reduces to a problem to find these values.

    The initial particles in our case were supposed to be nonrelativistic
at infinity: $ \varepsilon_{1} = \varepsilon_{2}=1 $,
so Eq.~(\ref{I4K}) becomes
    \begin{eqnarray}
- m \left[ \sqrt{ \frac{2 M}{r^3} \,(a - L_1)^2 + \frac{2 M}{r} -
\frac{L_1^2}{r^2}} \right.
    \hspace{22pt} \nonumber    \\[2pt]
+ \left. \sqrt{ \frac{2 M}{r^3} \,(a - L_2)^2 +
\frac{2 M}{r} - \frac{L_2^2}{r^2}} \, \right]
    \hspace{22pt} \nonumber    \\[2pt]
= \mu \left[ \sqrt{ \varepsilon_{1 \mu}^2
+\frac{2 M}{r^3} \,(a \varepsilon_{1 \mu} - L_{1 \mu})^2 +
\frac{a^2 \varepsilon_{1 \mu}^2 - L_{1 \mu}^2}{r^2} - \frac{\Delta}{r^2}}
\right.
    \nonumber  \\[2pt]
-\left. \sqrt{ \varepsilon_{2 \mu}^2 +
\frac{2 M}{r^3} \, (a \varepsilon_{2 \mu} - L_{2 \mu})^2 +
\frac{a^2 \varepsilon_{2 \mu}^2 - L_{2 \mu}^2}{r^2} - \frac{\Delta}{r^2}}
\, \right].
\label{I4Ken11}
\end{eqnarray}

    Consider at first the case~$a=0$ of the nonrotating black hole.
    In this case the Eq.~(\ref{I4Ken11}) has the form
    \begin{eqnarray}
\hspace*{-13pt}
- m \left[ \sqrt{
\frac{2M}{r} - \frac{L_1^2}{r^2} \left( 1- \frac{2M}{r} \right)}
+ \sqrt{
\frac{2M}{r} - \frac{L_2^2}{r^2} \left( 1- \frac{2M}{r} \right)}
\, \right]
\nonumber \\[4pt]
\hspace*{-42pt}= \mu \left[ \sqrt{ \varepsilon_{1 \mu}^2 -
\left(1 + \frac{L_{1 \mu}^2}{r^2} \right)\!
\left( 1 - \frac{2M}{r} \right)} \right.
\hspace{22pt}
\nonumber \\[4pt]
\hspace*{-42pt} -\left.
\sqrt{ \varepsilon_{2 \mu}^2 - \left(1 + \frac{L_{2 \mu}^2}{r^2} \right)\!
\left( 1 - \frac{2M}{r} \right)} \, \right]\!. \hspace{22pt}
\label{I3}
\end{eqnarray}
    The maximal energy of collision is for
$L_1 =-L_2, \ |L_1| =|L_2|=4 M$ (see~(\ref{geodKerr5}), (\ref{KerrLimAMax})).
    From Eq.~(\ref{I3}) one gets for $|L_{1 \mu}|=|L_{2 \mu}|$
(in particular, for the radial movement of the products of reaction)
$ \varepsilon_{1 \mu} \le \varepsilon_{2 \mu} $.
    The initial particles were supposed to be nonrelativistic on
infinity and so from~(\ref{I5K}) one gets for the energy of the particle
moving outside the black hole
    \begin{equation} \label{I4}
\mu \varepsilon_{1 \mu} \le m \,.
\end{equation}
    For radial movement the energy radiated to infinity of the particles
collided in the Schwarzschild black hole can not be larger than the rest
energy of one initial particle!

    Note that equality in~(\ref{I4}) is obtained in case
$L_1=-L_2$, \ $|L_1|=|L_2|= 4 M $, \
$L_{1 \mu} = L_{2 \mu} = 0 $ (the radial movement),
if $ r = 4 M $ then reaction occurs on the doubled Schwarzschild radius.

    For the case when the collision takes place on the horizon of the
black hole ($r \to r_H$) the system~(\ref{I5K})--(\ref{I4Ken11}) can be solved
exactly
    \begin{equation} \label{Inew}
\hspace*{-11pt}
\varepsilon_{1 \mu} \!=\! \frac{A L_{1 \mu}}{2r_H}, \
\varepsilon_{2 \mu} \!=\! \frac{2 m}{\mu} - \frac{A L_{1 \mu}}{2 r_H}, \
L_{2 \mu} \!=\! \frac{m}{\mu} (L_1+L_2) - L_{1 \mu} . \ \
\end{equation}

    In general case the system of three Eqs.~(\ref{I5K})--(\ref{I4Ken11})
for four variables
$\varepsilon_{1 \mu}, \ \varepsilon_{2 \mu}, \ L_{1 \mu}, \ L_{2 \mu}$
can be solved numerically for a fixed value of one variable
(and fixed parameters $m/\mu, \ L_1/M, \ L_2/M, \ a/M, \ r/M$).
    The example of numerical solution is
$\mu/m=0.3, l_1=2.2, \, l_2=2.198, \, A=0.99, \, x=1.21, \, l_{1 \mu} = 16.35,
\, l_{2 \mu} = -1.69, \, \varepsilon_{1 \mu}=7.215 ,  
\, \varepsilon_{2 \mu}=-0.548 $.  
    Note that the energy of the second final particle is negative and the
energy of the first final particle contrary to the limit obtained
in~\cite{JacobsonSotiriou09} is larger than the energy of initial particles as
it must be in the case of a Penrose process~\cite{PiranShahamKatz75}.
    If the mass of the particle is not very large this particle is observed
as ultrarelativistic.
    What is the reason of this contradiction?
    Let us investigate the problem carefully.

    Note that if one neglects the states with negative energy in ergosphere
energy extracted in the considered process can not be larger than
the initial energy of the pair of particles at infinity, i.e.~$2m$.
    The same limit~$2m$ for the extracted energy for any (including Penrose
process) scattering process in the vicinity of the black hole was
obtained in~\cite{JacobsonSotiriou09}.
    Let us show why this conclusion is incorrect.

    If the angular momentum of the falling particles is the same
then (see~(\ref{I4Ken11})) one has the situation similar to the usual
decay of the particle with mass~$2m$ in two particles with mass~$\mu$.
    Due to the Penrose process in ergosphere it is possible that the particle
falling inside the black hole has the negative relative to infinity energy
and then the extracted particle can have energy larger than~$2m$.

    The main assumption made in~\cite{JacobsonSotiriou09} is the supposition
of the collinearity of vectors of 4-momenta of the particles falling inside
and outside of the black hole
(see (9)--(11) in~\cite{JacobsonSotiriou09}).
    The authors of~\cite{JacobsonSotiriou09} say that these vectors are
``asymptotically tangent to the horizon generator''.

    First note that from~(\ref{geodKerr3}), (\ref{geodKerr5}) for $A<1$ and
$l \le l_R$ or $A=1 $, but $l<2$ the limit of $dr/d \tau $ is not zero at the
horizon.
    This derivative has opposite signs for the falling and outgoing particles.
    Signs for other components of the 4-momentum are equal.

    In the limiting case ($A=1$, $l_1=2$)
the expressions $d t / d \tau $,  $d \varphi / d \tau $
of the components of the 4-velocity of the infalling
particle~(\ref{geodKerr1}), (\ref{geodKerr2}) go to infinity when
$r \to r_H $, but $d r / d \tau $ goes to zero.
    In spite of smallness of $r$-components in the expression of the square
of the 4-momentum vector they have the factor $g_{rr} $ going to infinity
at the horizon.
    So putting them to zero can lead to a mistake.
    To see if $u_{(1)}$ and $v_{(1)}$ are collinear it is necessary to put
the coordinate~$r$ of the collision point to the limit $r_H=M$ and resolve
the uncertainties $\infty / \infty$ and $0/0$.
    For the falling particle $\varepsilon =1$, $l=2$ the expressions for
components of the 4-vector~$u$ can be easily found
from~(\ref{geodKerr1})--(\ref{geodKerr3}).
    For the particle outgoing from the black hole due to exact solution
on the horizon~(\ref{Inew}) one puts
$\varepsilon_{1 \mu} = l_{1 \mu}/2 + \alpha$, where $\alpha$ is some function
of~$r$ and $l_{1 \mu}$, such that $\alpha \to 0$ when $r \to r_H$.
    Putting this $\varepsilon_{1 \mu} $ into~(\ref{geodKerr1})--(\ref{geodKerr3})
one gets for $x=r/M \to 1$
    \begin{equation} \label{Otn1}
\frac{v^t_{(1)}}{u^t_{(1)}} = \frac{v^\varphi_{(1)}}{u^\varphi_{(1)}} =
\frac{\alpha}{x-1} + \frac{l_{1 \mu}}{2} \,,
\end{equation}
    \begin{equation} \label{Otn2}
\frac{v^r_{(1)}}{u^r_{(1)}} = - \sqrt{
\frac{ 2 \alpha^2 }{(x-1)^2} + \frac{ 2 l_{1 \mu} \alpha}{x-1} +
\frac{3}{8}\, l_{1 \mu}^2 - \frac{1}{2} } \,.
\end{equation}
    Due to the condition $dt / d \tau >0 $ (movement forward in time)
the necessary condition for collinearity is that
both~(\ref{Otn1}) and (\ref{Otn2})
must be zero, which is not true.
    This leads to the conclusion that the considerations of the authors
of~\cite{JacobsonSotiriou09} for scattering exactly on the horizon can not be
used for the real situation of particle scattering close to the horizon.

    All this confirms our hypothesis in~\cite{GribPavlov2007AGN}
that the vicinity of the rotating black hole is the arena of the Grand
Unification physics as it was in the early Universe.
    In the early Universe gravitation created pairs of superheavy
$X, \tilde{X}$ particles decaying then as short living and long living
$ X_S,\, X_L$ particles on ordinary quarks and leptons.
    But then due to the breaking of the GU symmetry $X_L$ became
metastable and survived up to our time as dark matter particles.

    In the vicinity of rotating black holes $X_L$ can decay on ordinary
particles as it is possible for the GU energies as well as they can be
converted into $X_S$ decaying particles because
$\langle X_L | X_S\rangle \ne 0 $.
    These decays can go with baryon and $CP$-conservation breaking,
so that different hypotheses on the processes in the early Universe can
be checked by astrophysical observations of AGN.

\section*{Acknowledgments}

The authors are indebted to Ted Jacobson and Thomas P. Sotiriou for
putting their attention to a numerical mistake in the first
version of the paper~\cite{GrPv10}.


\end{document}